\noindent\centerline{\bf Computable Solutions of Fractional Partial Differential}

\vskip.1cm\noindent\centerline{\bf Equations Related to Reaction-Diffusion Systems}

\vskip.3cm\noindent\centerline{R.K. Saxena}
\vskip.2cm\centerline{Department of Mathematics and Statistics}
\vskip0cm\centerline{JNV University, Jodhpur-342004, India}

\vskip.3cm\centerline{A.M. Mathai}
\vskip.2cm\centerline{Centre for Mathematical Sciences Pala, Kerala-686574, India}
\vskip0cm\centerline{and McGill University, Canada}

\vskip.3cm \centerline{H.J. Haubold}
\vskip.2cm\centerline{Office for Outer Space Affairs, United Nations,}
\vskip0cm\centerline{Vienna International Centre, 1400-Vienna, Austria}
\vskip0cm\centerline{and Centre for Mathematical Sciences Pala, Kerala-686574, India}

\vskip.3cm \noindent\centerline{ {\bf Abstract}}

\vskip.3cm

The object of this paper is to present a computable solution of a  fractional partial differential equation associated with a  Riemann-Liouville  derivative of fractional order as the  time-derivative  and Riesz-Feller fractional  derivative as the  space derivative. The method followed in deriving the solution is that of joint Laplace and Fourier transforms. The solution is derived in a closed and computable form in terms of the H-function. It provides an elegant extension of the  results given earlier by Debnath [4], Chen et al. [2],  Haubold et al. [17], Mainardi et al. [27,28 ], Saxena et al. [39], and Pagnini et al. [35].  The results obtained are presented in the form of four theorems. Some results associated with fractional Schr\"odinger equation and fractional diffusion-wave equation are also derived as special cases of the findings.


\vskip.3cm\noindent
{\bf 1.\hskip.3cm Introduction}

\vskip.3cm
General models for reaction-diffusion systems are  discussed by Henry and Wearne [18, 19], Henry et al. [20], Haubold et al. [16,17], Saxena et al. [38, 39, 40], Mainardi et al. [27, 28] and others. Stability in reaction-diffusion systems and  nonlinear oscillation have been discussed by Gafiychuk et al. [12,13]. Recently, Engler [7] obtained the speed of spread for fractional reaction-diffusion. Distributed order sub-diffusion is discussed by Naber [32]. In a recent paper, Chen et al. [2] have derived the fundamental and numerical solution of a reaction diffusion equation associated with the Riesz fractional derivative as the space derivative. Reaction-diffusion models associated with Riemann-Liouville fractional derivative as the time fractional derivative and Riesz-Feller derivative as the space derivative are recently discussed by Haubold et al. [17]. Such equations in case of Caputo fractional derivative are also solved by Saxena et al. [39]. In connection with the evolution equations for the probabilistic generalization of the Voigt profile function, it is shown by Pagnini and Mainardi [35] that the solution of the following integro-differential equation

$${{\partial N}\over{\partial \tau}}={_0D}_0^{\alpha_1}N(x,t)+{_0D}_0^{\alpha_2}N(x,t),~~N(x,0)=\delta(x)\eqno(1.1)
$$in terms of its Fourier transform, where ${_0D}_0^{\alpha_1}$ and ${_0D}_0^{\alpha_2}$ are the Riesz fractional derivatives of orders $\alpha_1$ and $\alpha_2$ respectively, and $\delta(x)$   is the Dirac-delta function, which is given in [35, p.1593]. Consider the following Fourier transform, with the Fourier parameter $k$:

$$N^{*}(k,\tau)=\exp\{-\tau(|k|^{\alpha_1}+|k|^{\alpha_2})\}.\eqno(1.2)
$$This has motivated the authors to investigate the solutions of fractional partial differential  equations (3.1)  and  (3.14). The technique used in deriving the results is that of joint Laplace and Fourier transforms. The results are obtained in a closed and computable form. Due to the general character of the derived results, many known results given earlier by Chen et al. [2], Haubold et al. [17], Saxena et al. [39], Pagnini et al. [35], and others readily follow as special cases of our derived results. Solutions of certain extended space-time fractional diffusion wave equation and  generalized Schr\"odinger equations are also deduced from our findings.\vskip.2cm
 	The Schr\"odinger equation is a fundamental equation of quantum mechanics. Feynman and Hibbs [11] reconstructed the Schr\"odinger equation by making use of the path integral approach by considering a Gaussian probability distribution. This approach is further extended by Laskin [23-25] in formulating the fractional Schr\"odinger equation by generalizing the Feynman path integrals from Brownian-like to Levy-like quantum mechanical paths.  In a similar manner, one obtains a time fractional equation if non-Markovian evolution is considered. In a recent paper, Naber [32] discussed certain properties of the time fractional Schr\"odinger equation by writing the Schr\"odinger equation in terms of  fractional derivatives as dimensionless variables. The time fractional Schr\"odinger equations are  discussed by  Debnath [4,5], Bhatti and Debnath [1], Saxena et al. [38], Tofight [40], and others.

\vskip.3cm\noindent
{\bf 2.\hskip.3cm Mathematical Prerequisites}
\vskip.3cm

The Riemann-Liouville fractional derivative of order $\alpha >0$ is defined as (Samko et al. [37, p.37]; also see, Kilbas et al. [21])

  $${_0D}_t^{\alpha}f(x,t)={{1}\over{\Gamma(n-\alpha)}}{{{\rm d}^n}\over{{\rm d}t^n}}\int_0^t(t-\tau)^{n-\alpha-1}f(x,t){\rm d}t,~n=[\alpha]+1,~n\in N,~t>0,\eqno(2.1)
  $$where $[\alpha]$  means the integer part of the number $\alpha$. The Laplace transform of the  Riemann-Liouville fractional derivative  is given by Oldham and Spanier [34, eq.(3.1.3)]; (also see, Kilbas et al.  [21]):
 	
$$L\{{_0D}_t^{\alpha}N(x,t);s\}=s^{\alpha}\tilde{N}(x,s)-\sum_{r=1}^ns^{r-1}{_0D}_t^{\alpha-r}N(x,t)|_{t=0},~n-1<\alpha\le n.\eqno(2.2)
$$This derivative is useful in deriving the solutions of  integral equations of fractional order governing certain physical problems of anomalous reaction and anomalous diffusion. In this connection, one can refer to the monograph by Podlubny [36], Samko et al. [37], Oldham and Spanier [34], Miller and Ross [31], Kilbas et al. [21], Mainardi [26], Diethelm [6], and recent papers on the subject  [27,28,30,33,40,41].
 Following Feller [9,10], it is conventional  to define  the Riesz-Feller space-fractional derivative of order $\alpha$  and skewness $\theta$   in terms of its Fourier transform as

  $$\eqalignno{F\{{_xD}_{\theta}^{\alpha}f(x);k\}&=-\Psi_{\alpha}^{\theta}(k)f^{*}(k),&(2.3)\cr                                                              		 \noalign{\hbox{where}}
   \Psi_{\alpha}^{\theta}(k)&=|k|^{\alpha}\exp[i(sign k){{\theta\pi}\over{2}}],~0<\alpha\le 2,~|\theta|\le\min\{\alpha,2-\alpha\}.&(2.4)\cr}
   $$Further when $\theta=0$, we have a symmetric operator with respect to $x$ that can be interpreted as

$${_xD}_0^{\alpha}=-\left(-{{{\rm d}^2}\over{{\rm d}x^2}}\right)^{{\alpha}\over2}.\eqno(2.5)
$$This can be formally deduced by writing $-(k)^{\alpha}=-(k^2)^{{\alpha}\over2}$. For $0<\alpha<2$ and $|\theta|\le\min\{\alpha,2-\alpha\}$, the Riesz-Feller derivative can be shown to possess the following integral representation in $x$ domain:                                                                                   		 
$$\eqalignno{{_xD}_{\theta}^{\alpha}f(x)&={{\Gamma(1+\alpha)}\over{\pi}}
\bigg\{\sin[(\alpha+\theta)\pi/2]\int_0^{\infty}{{f(x+\zeta)-f(x)}\over{\zeta^{1+\alpha}}}
{\rm d}\zeta\cr
&+\sin[(\alpha-\theta)\pi/2]\int_0^{\infty}{{f(x-\zeta)-f(x)}\over{\zeta^{1+\alpha}}}
{\rm d}\zeta\bigg\}&(2.6)\cr}
$$
For $\theta=0$, the Riesz-Feller fractional derivative becomes the  Riesz fractional derivative of order $\alpha$ for $1<\alpha\le 2$,   defined by analytic continuation in the whole range $0<\alpha\le 2,~\alpha\ne 1$ (see, Gorenflo and Mainardi [14]) as

$$\eqalignno{{_xD}_0^{\alpha}&=-\lambda[I_{+}^{-\alpha}-I_{-}^{-\alpha}],&(2.7)\cr	 							         \noalign{\hbox{where}}
\lambda&= {{1}\over{2\cos(\alpha\pi/2)}};~~I_{+}^{-\alpha}={{{\rm d}^2}\over{{\rm d}x^2}}I_{+}^{2-\alpha}.&(2.8)\cr}
$$The Weyl fractional integral operators  are defined in the monograph  by  Samko et al. [37]  as

$$\eqalignno{(I_{+}^{\beta}N)(x)&={{1}\over{\Gamma(\beta)}}\int_{-\infty}^{\infty}
(x-\zeta)^{\beta-1}N(\zeta){\rm d}\zeta,~\beta>0\cr
\noalign{\hbox{and}}
(I_{-}^{\beta}N)(x)&={{1}\over{\Gamma(\beta)}}\int_x^{\infty}(\zeta-x)^{\beta-1}
N(\zeta){\rm d}\zeta,~\beta>0.\cr}
$$					
\noindent {\bf Note 1}.\hskip.3cm We note that ${_xD}_0^{\alpha}$    is a pseudo  differential  operator. In particular
we have
  $$\eqalignno{{_xD}_0^2&={{{\rm d}^2}\over{{\rm d}x^2}},\hbox{  but  }{_xD}_0^1\ne {{{\rm d}}\over{{\rm d}x}}.&(2.10)\cr                                                  \noalign{\hbox{For $\theta=0$, we  have}}
  F\{{_xD}_0^{\alpha}f(x);k\}&=-|k|^{\alpha}f^{*}(k).&(2.11)\cr}
 $$The H-function is defined by means of a Mellin-Barnes type integral in the following manner [29, p.2]:
$$\eqalignno{H_{p,q}^{m,n}(z)&=H_{p,q}^{m,n}\left[x\big\vert_{(b_q,B_q)}^{(a_p,A_p)}
\right]\cr
&=H_{p,q}^{m,n}\left[x\big\vert_{(b_1,B_1),...,(b_q,B_q)}^{(a_1,A_1),...,(a_p,A_p)}\right]
={{1}\over{2\pi i}}\int_{\Omega}\Theta(\xi)z^{-\xi}{\rm d}\xi,&(2.12)\cr
\noalign{\hbox{where $i=(-1)^{1\over2}$,}}
 \Theta(\xi)&={{\{\prod_{j=1}^m\Gamma(b_j+B_j\xi)\}\{\prod_{j=1}^n\Gamma(1-a_j-A_j\xi)\}}
 \over{\{\prod_{j=m+1}^q\Gamma(1-b_j-B_j\xi)\}\{\prod_{j=n+1}^p\Gamma(a_j+A_j\xi)\}}},&(2.13)\cr}
 $$and an empty product is always interpreted as unity ; $m,~ n,~ p,~ q\in N_0$   with
$0\le   n\le   p,~ 1\le m\le  q,$ $A_j,R_{+},~a_j,b_j\in R$ or $ C$,~ $i=1...,p$;~ $j=1,...,q)$ such that
    $$A_i(b_j+k)\ne B_j(a_i-s-1),~k,s\in N_0~i=1,...,n;~ j=1,...,m,\eqno(2.14)                  $$and these poles are separated, where we employ the usual notations:  $N_0=(0,1,2,...)$;~ $R = (-\infty,\infty)$, and $C$ being the complex number field. A comprehensive account of the H-function is available from the monographs  Mathai et al. [29] and Kilbas et al. [22].
We also need the following result in the analysis that follows;
Haubold  et al. [16] has shown that
	 $$F^{-1}[E_{\beta,\gamma}(-at^{\beta}\Psi_{\alpha}^{\theta}(k);x]={{1}\over{\alpha|x|}}H_{3,3}^{2,1}\left[{{|x|}\over{(at^{\beta})^{1\over{\alpha}}}}\bigg\vert_{(1,{{1}\over{\alpha}}),(1,1),(1,\rho)}^{(1,{{1}\over{\alpha}}),(\gamma,{{\beta}\over{\alpha}}), (1,\rho)}\right],\eqno(2.15)
$$where $\Re(\alpha)>0,~\Re(\beta)>0,~\Re(\gamma)>0$.

\vskip.3cm\noindent{\bf 3.\hskip.3cm Solution of Unified Fractional Partial Differential Equations}

\vskip.3cm
In this section, we will investigate the solution of  fractional partial  differential equations, which may be regarded  as an extension of one-dimensional fractional reaction-diffusion equation, one-dimensional space-time fractional diffusion-wave equation and one-dimensional fractional Schr\"odinger equation. The results are presented in the form of the following four theorems.

\vskip.3cm\noindent{\bf
Theorem 1.}\hskip.3cm{\it Consider the following one-dimensional  non-homogeneous  unified fractional differential equation:

$${_0D}_t^{\alpha}N(x,t)=\lambda~{_xD}_{\theta}^{\beta}N(x,t)+\mu~{_xD}_{\phi}^{\gamma}U(x,t),\eqno(3.1)
$$where $t>0,~x\in R;~\alpha,~\theta,~\beta,~\gamma$  and $\phi$  are real parameters with the constraints

$$0<\beta\le 2,~0<\gamma\le 2,~|\theta|\le\min(\beta,2-\beta),~|\phi|\le\min(\gamma,2-\gamma),~0<\alpha\le 1,\eqno(3.2)
$$with the initial conditions

$${_0D}_t^{\alpha-1}N(x,0)=f(x),\hbox{  for  }x\in R,~\lim_{x\rightarrow \pm\infty}N(x,t)=0,~t>0.\eqno(3.3) 	
$$Here ${_0D}_t^{\alpha-1}N(x,0)$ means the Riemann-Liouville fractional partial  derivative of  $N(x, t)$ with respect to $t$ of order $\alpha-1$ evaluated at $t=0$; ${_xD}_{\theta}^{\beta}$   and ${_xD}_{\phi}^{\gamma}$  are the Riesz-Feller space-fractional derivatives  respectively of orders $\beta$  and $\gamma$  with asymmetries $\theta$  and $\phi$. ${_0D}_t^{\alpha}$ is the Riemann-Liouville time-fractional  derivative  of order $\alpha$;  $\lambda$  and $\mu$  are arbitrary constants, and $f(x)$  and $U= U(x,t)$ are the given functions. Then for the solution of (3.1), subject to the above conditions, there holds the formula

 $$\eqalignno{N(x,t)&= \int_{-\infty}^{\infty}G(x-\xi,t)f(\xi){\rm d}\xi\cr
 &-\mu\int_0^t(t-\tau)^{\alpha-1}[\int_{-\infty}^{\infty}G_1(x-\tau,t-\tau)
 U(\xi,\tau){\rm d}\xi]{\rm d}\tau, &(3.4)\cr
\noalign{\hbox{where the Green functions  $G(x,t)$ and $G_1(x,t)$  are given by}}
G(x,t)&={{t^{\alpha-1}}\over{2\pi}}\int_{-\infty}^{\infty}\exp(-ikx)
E_{\alpha,\alpha}(-\lambda\Psi_{\beta}^{\theta}(k)t^{\alpha}){\rm d}k,&(3.5)\cr
&={{t^{\alpha-1}}\over{\beta|x|}}H_{3,3}^{2,1}\left[{{|x|}\over{(\lambda~
t^{\alpha})^{1\over{\beta}}}}\bigg\vert_{(1,1),(1,{{1}\over{\beta}}),
(1,\rho)}^{(1,{{1}\over{\beta}}),(\alpha,{{\alpha}\over{\beta}}),(1,\rho)}\right]&(3.6)\cr
\noalign{\hbox{with $\beta>0,~\rho={{\beta-\theta}\over{2\beta}}$, and}}
G_1(x,t)&={{1}\over{2\pi}}\int_{-\infty}^{\infty}\exp(-ikx)\Psi_{\gamma}^{\phi}(k)
E_{\alpha,\alpha}(-\lambda~
\Psi_{\beta}^{\theta}(k)
t^{\alpha}){\rm d}k&(3.7)\cr}
$$and $H_{3,3}^{2,2}(z)$  is the H-function defined by (2.12) and $E_{\alpha,\beta}(z)$ is the Mittag-Leffler function, defined by  [8, Section 18.1]

$$E_{\alpha,\beta}(z)=\sum_{n=0}^{\infty}{{z^n}\over{\Gamma(\alpha n+\beta)}},~\alpha,~\beta\in C,~\Re(\alpha)>0,~\Re(\beta)>0.\eqno(3.8)	
$$}

\vskip.3cm In deriving the value of the integral in (3.5), the formula (2.15) has been used.

\vskip.3cm\noindent{\bf Proof:}\hskip.3cm In order to derive the solution of (3.1), we introduce the joint Laplace-Fourier transform in the form

$${\tilde{N}}^{*}(k,s)=\int_0^{\infty}\int_{-\infty}^{\infty}{\rm e}^{-st+ikx}N(x,t){\rm d}x~{\rm d}t,\eqno(3.9)	
$$where $\Re(s)>0,~k>0$.  If we apply the Laplace transform with respect to the time variable $t$, Fourier transform with respect to space variable $x$ and use the initial conditions (3.2), (3.3) and the formula (2.2), then the given equation transforms into the form

$$s^{\alpha}{\tilde{N}}^{*}(k,s)-f^{*}(k)
=-\lambda~\Psi_{\alpha}^{\theta}(k){\tilde{N}}^{*}(k,s)
-\mu~\Psi_{\gamma}^{\phi}(k){\tilde{U}}^{*}(k,s),     $$where according to the conventions followed, the symbol ${\tilde{N}(k,s)} $ will stand for the Laplace transform with respect to time variable $t$ and  $*$  represents the Fourier transform with respect to space variable $x$. Solving for ${\tilde{N}}^{*}(k,s)$, it yields

$${\tilde {N}}^{*}(k,s)={{f^{*}(k)}\over{s^{\alpha}+\lambda~\Psi_{\beta}^{\theta}(k)}}
-{{\mu~\Psi_{\gamma}^{\phi}(k){\tilde{U}}^{*}(k,s)}\over{s^{\alpha}
+\lambda~\Psi_{\beta}^{\theta}(k)}}.\eqno(3.10)
$$To invert (3.10), it is convenient to first invert the Laplace transform and then the Fourier transform.  Thus to invert the Laplace transform we use the formula given in [39]

 $$L^{-1}\{{{s^{\beta-1}}\over{a+s^{\alpha}}};t\}=t^{\alpha-\beta}E_{\alpha,\alpha-\beta+1}(-at^{\alpha}),\eqno(3.11)
$$where $\Re(s) > 0,~\Re(\alpha)>0,~\Re(\alpha-\beta)>-1$ and the convolution theorem of the Laplace transform to obtain

 $$\eqalignno{N^{*}(k,t)&=f^{*}(k)t^{\alpha-1}E_{\alpha,\alpha}(-\lambda~t^{\alpha})\Psi_{\beta}^{\theta}(k)\cr
 &-t^{\mu}~\Psi_{\gamma}^{\phi}(k)\int_0^t(t-\tau)^{\alpha-1}E_{\alpha,\alpha}
 (-\alpha\Psi_{\beta}^{\theta}(k)(t-\tau)^{\alpha})U^{*}(k,\tau){\rm d}\tau.&(3.12)\cr}
$$Now the application of the inverse Fourier transform gives the exact solution in the following form:

$$\eqalignno{N(k,t)&=t^{\alpha-1}F^{-1}[f^{*}(k)E_{\alpha,\alpha}
(-\lambda~\Psi_{\gamma}^{\phi}(k)t^{\alpha})]\cr
&-\mu\int_0^t(t-\tau)^{\alpha-1}F^{-1}[\Psi_{\gamma}^{\phi}(k)
E_{\alpha,\alpha}(-\lambda~\Psi_{\beta}^{\theta}(k)(t-\tau)^{\alpha})U^{*}(k,\tau)]{\rm d}\tau.&(3.13)\cr}
$$Finally, the application of the convolution theorem of the Fourier transform yields the  desired solution (3.4).\vskip.2cm

Following a similar procedure, it is not difficult to establish the following theorems:

\vskip.3cm\noindent{\bf
Theorem 2.}\hskip.3cm{\it  Consider the same equation in (3.1) with the same condition on the parameters except that $0<\alpha\le 2$, instead of $0<\alpha\le 1$. In addition to the initial conditions in (3.3), assume that

 $${_0D}_t^{\alpha-2}N(x,0)=g(x).\eqno(3.14)
 $$Then the solution of (3.1) under the conditions (3.2), (3.3) and (3.14) is given by

 $$\eqalignno{N(x,t)&=\int_{-\infty}^{\infty}G(x-\xi,t)f(\xi){\rm d}\xi+\int_{-\infty}^{\infty}G_2(x-\xi,t)g(\xi){\rm d}\xi\cr
 &-\mu\int_0^t(t-\tau)^{\alpha-1}[\int_{-\infty}^{\infty}G_1(x-\tau,t-\tau)U(\xi,\tau){\rm d}\xi]{\rm d}\tau,&(3.15)\cr}
$$where the Green functions  $G(x,t)$ and $G_1(x,t)$  are  defined in (3.6)  and (3.7) respectively and $G_2(x,t)$   is given by

$$\eqalignno{G_2(x,t)&={{t^{\alpha-2}}\over{2\pi}}\int_{-\infty}^{\infty}\exp(-ikx)
E_{\alpha,\alpha-1}(-\lambda~\Psi_{\beta}^{\theta}(k)t^{\alpha}){\rm d}k\cr
&={{t^{\alpha-2}}\over{\beta|x|}}H_{3,3}^{2,1}\left[{{|x|}\over{(\lambda t^{\alpha})^{1\over{\beta}}}}\bigg\vert_{(1,1),(1,{{1}\over{\beta}}),
(1,\rho)}^{(1,{{1}\over{\beta}}),(\alpha-1,{{\alpha}\over{\beta}}), (1,\rho)}\right],~\beta>0&(3.16)\cr}
$$with $\rho={{\beta-\theta}\over{2\beta}}$ and $H_{3,3}^{2,1}(z)$  is the H-function defined by (2.17) and $E_{\alpha,\beta}(z)$  is the Mittag-Leffler function, defined in (3.8).}

\vskip.3cm

Now, we set $U(x,t) =N(x,t)$, where $N(x,t)$ is the unknown function to arrive at the next result.

\vskip.3cm\noindent{\bf Theorem 3.}\hskip.3cm{\it Consider the same equation in (3.1) with $U(x,t)$ replaced by $N(x,t)$. Assume that the conditions (3.2) and (3.3) hold. Then solution of (3.1) with $U(x,t)$ replaced by $N(x,t)$ is given by

$$\eqalignno{N(x,t)&=\int_{-\infty}^{\infty}G_3(x-\xi,t)f(\xi){\rm d}\xi,&(3.17)\cr
\noalign{\hbox{where the Green function $G_3(x,t)$  is given by}}
 G_3(x,t)&={{t^{\alpha-1}}\over{2\pi}}\int_{-\infty}^{\infty}\exp(-ikx)
 E_{\alpha,\alpha}[-(\lambda~\Psi_{\beta}^{\theta}(k)+\mu~\Psi_{\gamma}^{\theta}(k))
 t^{\alpha}]{\rm d}t.&(3.18)\cr}$$}

 \vskip.3cm
Similarly, we find that the following theorem holds true:

\vskip.3cm\noindent{\bf Theorem 4.}\hskip.3cm{\it  Consider the same equation in Theorem 3 with $0<\alpha\le 2$ and under the conditions (3.2), (3.3), and (3.14). Then the solution of (3.1) with $U(x,t)$ replaced by $N(x,t)$ with $0<\alpha\le 2$ is given by

 $$N(x,t)=\int_{-\infty}^{\infty}G_3(x-\xi,t)f(\xi){\rm d}\xi,\eqno(3.19)
$$where the Green function $G_3(x,t)$  is defined in  (3.18) and the other Green function $G_4(x,t)$ is given by

$$G_4(x,t)={{t^{\alpha-2}}\over{2\pi}}\int_{-\infty}^{\infty}\exp(-ikx)E_{\alpha,\alpha-1}
[-(\lambda~\Psi_{\beta}^{\theta}(k)+\mu~\Psi_{\gamma}^{\psi}(k))t^{\alpha}]{\rm d}k.\eqno(3.20)$$}

\vskip.3cm\noindent
 {\bf Note 2}.\hskip.3cm It is interesting to observe that for $g(x) =0$, Theorems 2 and 4 yield similar types of results as Theorems  1 and 3 respectively.
 \vskip.3cm\noindent{\bf 4.\hskip.3cm Selected Special Cases}

 \vskip.3cm

If we set $\theta=\phi=0$  then by virtue of the identity (2.11), Riesz-Feller derivative reduces to Riesz derivative and consequently Theorem 1 yields the following results:

\vskip.3cm\noindent{\bf Corollary 1.}\hskip.3cm{\it Consider the one-dimensional non-homogeneous unified fractional differential equation (3.1) for $\theta=\phi=0$:

$${_0D}_t^{\alpha}N(x,t)=\lambda~{_xD}_0^{\beta}N(x,t)+\mu~{_xD}_0^{\gamma}U(x,t),\eqno(4.1)
$$where the conditions (3.2) and (3.3) hold under $\theta=\phi=0$, $t>0,~x\in R,~\alpha,~\beta,~\gamma$ are real. Then for the solution of (4.1) there holds the formula

$$\eqalignno{N(x,t)&=\int_{-\infty}^{\infty}G_5(x-\xi,t)f(\xi){\rm d}\xi\cr
&-\mu\int_0^t(t-\tau)^{\alpha-1}F^{-1}[\int_{-\infty}^{\infty}G_6(x-\tau,t-\tau) U(\xi,\tau){\rm d}\xi]{\rm d}\tau,&(4.2)\cr
\noalign{\hbox{where the Green functions $G_5(x,t)$  and $G_6(x,t)$  are given by}}
G_5(x,t)&={{t^{\alpha-1}}\over{2\pi}}\int_{-\infty}^{\infty}\exp(-ikx)E_{\alpha,\alpha}(-\lambda|k|^{\beta}t^{\alpha}){\rm d}k&(4.3)\cr
&={{1}\over{\beta|x|}}H_{3,3}^{2,1}\left[{{|x|}\over{(\lambda t^{\alpha})^{1\over{\beta}}}}\bigg\vert_{(1,1),(1,{{1}\over{\beta}}),(1,{1\over2})}^{(1,{{1}\over{\beta}}),(\alpha,{{\alpha}\over{\beta}}),(1,{1\over2})}\right],~\beta>0&(4.4)\cr
G_6(x,t)&={{1}\over{2\pi}}\int_{-\infty}^{\infty}\exp(-ikx)|k|^{\gamma}E_{\alpha,1}
(-\lambda|k|^{\beta} t^{\alpha}){\rm d}k.&(4.5)\cr}$$}
	 				
 \vskip.3cm

For $\mu=\phi=0$,  Theorem 1  reduces to the following result given by Haubold et al. [10].

\vskip.3cm\noindent{\bf Corollary 2.}\hskip.3cm{\it Consider the following one-dimensional  fractional reaction-diffusion model of (3.1) with $\mu=\phi=0$

$${_0D}_t^{\alpha}N(x,t)=\lambda~{_xD}_{\theta}^{\beta}N(x,t),\eqno(4.6)
$$where $t>0,~x\in R,~\alpha,~\theta,~\beta$ real, and conditions (3.2) and (3.3) hold under $\mu=\phi=0$. Then for the solution of (4.6) there holds the formula

$$N(x,t)=\int_{-\infty}^{\infty}G(x-\xi,t)f(\xi){\rm d}\xi,\eqno(4.7)
$$where the Green function  $G(x,t)$ is  defined in (3.6).}

\vskip.3cm
If we further set $f(x)=\delta(x)$, where $\delta(x)$ is the Dirac-delta function, we obtain the fundamental solution of the  space-time fractional diffusion equation given by  Haubold et al. [17].\vskip.2cm

	By setting $\theta=\phi=0$  and using (2.11), the following  Corollaries 3-5  can be easily deduced from Theorems 2-4.

\vskip.3cm\noindent{\bf Corollary 3.}\hskip.3cm{\it  Consider the following one-dimensional  non-homogeneous  unified fractional differential equation:

$${_0D}_t^{\alpha}N(x,t)=\lambda~{_xD}_0^{\beta}N(x,t)+\mu~{_xD}_0^{\gamma}U(x,t),\eqno(4.8)
$$where (3.2), (3.3), (3.14) hold under $\theta=\phi=0$. Then for the solution of (4.8), there holds the formula

$$\eqalignno{N(x,t)&=\int_{-\infty}^{\infty}G_5(x-\xi,t)f(\xi){\rm d}\xi+\int_{-\infty}^{\infty}G_7(x-\xi,t)g(\xi){\rm d}\xi,\cr
&-\mu\int_0^t(t-\tau)^{\alpha-1}F^{-1}[\int_{-\infty}^{\infty}G_6(x-\tau,t-\tau)
U(\xi,\tau){\rm d}\xi]{\rm d}\tau,&(4.9)\cr}
$$where the Green functions $G_5(x,t)$ and $G_6(x,t)$ are defined in (4.3)  and (4.5) respectively and $G_7(x,t)$  is given by

$$\eqalignno{G_7(x,t)&={{t^{\alpha-2}}\over{2\pi}}\int_{-\infty}^{\infty}\exp(-ikx)E_{\alpha,\alpha}(-\lambda|k|^{\beta}t^{\alpha}){\rm d}k\cr
&={{t^{\alpha-2}}\over{\beta|x|}}H_{3,3}^{2,1}\left[{{|x|}\over{(\lambda t^{\alpha})^{1\over{\beta}}}}\bigg\vert_{(1,1),(1,{{1}\over{\beta}}),(1,{1\over2})}^{(1,{{1}\over{\beta}}),(\alpha-1,{{\alpha}\over{\beta}}),(1,{1\over2})}\right],~\beta>0,&(4.10)\cr}
 $$where $\rho={{\beta-\theta}\over{2\beta}}$.}

 \vskip.3cm\noindent{\bf Corollary 4.}\hskip.3cm{\it Consider the following one-dimensional  non-homogeneous  unified fractional differential equation
where $\alpha,~\beta$ and $\gamma$ are real parameters with the constraints

$${_0D}_t^{\alpha}N(x,t)=\lambda~{_xD}_0^{\beta}N(x,t)+\mu~{_xD}_0^{\gamma}N(x,t),\eqno(4.11)
$$where (3.2), (3.3) hold for $\theta=\phi=0$. Then

$$\eqalignno{N(x,t)&=\int_{-\infty}^{\infty}G_8(x-\xi,t)f(\xi){\rm d}\xi,&(4.12)\cr
\noalign{\hbox{where the Green function $G_8(x,t)$  is given by}}
G_8(x,t)&={{t^{\alpha-1}}\over{2\pi}}\int_{-\infty}^{\infty}\exp(-ikx)
E_{\alpha,\alpha}[-(\lambda|k|^{\beta}+\mu~|k|^{\gamma})t^{\alpha}]{\rm d}k.&(4.13)\cr}$$}

\vskip.3cm\noindent{\bf Corollary 5.}\hskip.3cm{\it Consider the following one-dimensional non-homogeneous  unified fractional differential equation

$${_0D}_t^{\alpha}N(x,t)=\lambda~{_xD}_0^{\beta}N(x,t)+\mu~{_xD}_0^{\gamma}N(x,t),\eqno(4.14)
$$where (3.2), (3.3), (3.14) hold. Then

$$N(x,t)=\int_{-\infty}^{\infty}G_8(x-\xi,t)f(\xi){\rm d}\xi+\int_{-\infty}^{\infty}t~G_9(x-\xi,t)g(\xi){\rm d}\xi,\eqno(4.15)
$$where the Green functions $G_8(x,t)$   is defined in (4.13) and the other Green function $G_9(x,t)$ is given by

 $$G_9(x,t)={{t^{\alpha-2}}\over{2\pi}}\int_{-\infty}^{\infty}\exp(-ikx)
 E_{\alpha,\alpha-1}[-(\lambda|k|^{\beta}+\mu|k|^{\gamma})t^{\alpha}]{\rm d}k.\eqno(4.16)$$}

 \vskip.3cm\noindent{\bf Note 3.}\hskip.3cm It is observed that for $g(x) =0$, Theorems 2 and 3 yield similar types of results as Theorems 1 and 3, respectively.
\vskip.3cm\noindent{\bf 5.\hskip.3cm  Further Special Cases}

\vskip.3cm
In this section we discuss  some consequences of the main results relating to the fractional Schr\"odinger equation. In the following corollaries 6 and 7, $h$ is the Planck constant.
\vskip.2cm\noindent
If we set $\gamma-\phi=0$  and $\lambda={{ih}\over{2m}}$, then Theorems 1 and 2 give the solutions of non-homogeneous  fractional   Schr\"odinger equations as shown  below.

\vskip.3cm\noindent{\bf Corollary 6.}\hskip.3cm{\it Consider the following one-dimensional non-homogeneous unified  fractional Schr\"odinger  equation of mass m:

   $${_0D}_t^{\alpha}N(x,t)=({{ih}\over{2m}}){_xD}_{\theta}^{\beta}N(x,t)+\mu~U(x,t),\eqno(5.1)                                                           $$where $t>0,~x\in R,~\alpha,~\beta$ real, with the constraints

     $$0<\beta\le 2,~0<\alpha\le 1,\eqno(5.2)
$$with the initial conditions (3.3), and $N(x,t)$ as the wave function.

 Then for the solution of (5.1), subject to the above constraints, there holds the formula
$$\eqalignno{N(x,t)&=\int_{-\infty}^{\infty}G(x-\xi,t)f(\xi){\rm d}\xi\cr
&+\int_{-\infty}^{\infty}(t-\xi)^{\alpha-1}[\int_{-\infty}^{\infty}G_{10}(x-\xi,t-\tau)
U(\xi,\tau){\rm d}\xi]{\rm d}\tau,&(5.3)\cr}
$$where the Green function  $G(x,t)$ is defined in (3.6) and $G_{10}(x,t)$ is given by
$$\eqalignno{
G_{10}(x,t)&=\int_{-\infty}^{\infty}\exp(-ikx)E_{\alpha,\alpha}[-(\lambda~
\Psi_{\beta}^{\theta}(k))
t^{\alpha}]{\rm d}k\cr
&={{1}\over{\beta|x|}}H_{3,3}^{2,1}\left[{{|x|}\over{(at^{\alpha})^{1\over{\beta}}}}
\bigg\vert_{(1,1),(1,{{1}\over{\beta}}),(1,\rho)}^{(1,{{1}\over{\beta}}),(\alpha,
{{\alpha}\over{\beta}}),(1,\rho)}\right],&(5.4)\cr}
$$where $\beta>0,~a={{ih}\over{2m}},~\rho={{\beta-\theta}\over{2\beta}}$.}

\vskip.3cm\noindent{\bf Corollary 7.}\hskip.3cm{\it Consider the following one-dimensional non-homogeneous  unified fractional Schr\"odinger  equation of a particle of mass m:

 $${_D}_t^{\alpha}N(x,t)=({{ih}\over{2m}}){_xD}_{\theta}^{\beta}N(x,t)+\mu~U(x,t),\eqno(5.5)
 $$where $t>0,~x\in R,~\alpha,~\theta,~\beta$ real,  with the constraints

$$0<\beta\le 2,~|\theta|\le\min(\beta,2-\beta),~1<\alpha\le 2,\eqno(5.6)
$$and with the initial conditions (3.3). Then for the solution of (5.5), subject to the above constraints, there holds the formula

$$\eqalignno{N(x,t)&=\int_{-\infty}^{\infty}G(x-\xi,t)f(\xi){\rm d}\xi+\int_{-\infty}^{\infty}t~G_2(x-\xi,t)g(\xi){\rm d}\xi\cr
&+\mu\int_0^t(t-\tau)^{\alpha-1}[\int_{-\infty}^{\infty}
G_{10}(x-\tau,t-\tau)U(\xi,\tau){\rm d}\xi]{\rm d}\tau,&(5.7)\cr}
$$where the Green functions  $G(x,t), G_2(x,t)$ and $G_{10}(x,t)$  are respectively given by (3.6), (3.16), and (5.4), respectively.}

\vskip.3cm

On the other hand, if we set $\gamma=\phi=0$, Theorems 1 and 2  provide the solutions of non-homogeneous fractional generalized diffusion-wave equations as given below:

\vskip.3cm\noindent{\bf Corollary 8.}\hskip.3cm{\it Consider the following one-dimensional non-homogeneous unified fractional diffusion-wave equation:

$${_0D}_t^{\alpha}N(x,t)=c^2~{_xD}_{\theta}^{\beta}N(x,t)+\mu~U(x,t),\eqno(5.8)
$$where $t>0,x\in R,~\alpha,~\beta$ real, with the constraints

$$0<\beta\le 2,~0<\alpha\le 1,\eqno(5.9)
$$and with the initial conditions (3.3) and (5.2), where $\mu$ and $c$ are arbitrary constants.
Then for the solution of (5.8), subject to the above conditions, there holds the formula

$$\eqalignno{N(x,t)&=\int_{-\infty}^{\infty}G(x-\xi,t)f(\xi){\rm d}\xi\cr
&+\int_{-\infty}^{\infty}(t-\xi)^{\alpha-1}[\int_{-\infty}^{\infty}G_{10}(x-\xi,t-\tau)U(\xi,\tau){\rm d}\xi]{\rm d}\tau,&(5.10)\cr}
$$where the Green function  $G(x,t)$ and $G_{10}(x,t)$ are respectively given in  (3.6)  and (5.4).}

\vskip.3cm\noindent{\bf Corollary 9.}\hskip.3cm{\it Consider the following one-dimensional non-homogeneous unified fractional diffusion-wave equation:

$${_0D}_t^{\alpha}N(x,t)=c^2~{_xD}_{\theta}^{\beta}N(x,t)+\mu~U(x,t),\eqno(5.11)
$$where $t>0,~x\in R,~\alpha,~\theta,~\beta$ real, with the constraints
$0<\beta\le 2$, $|\theta|\le \min(\beta,2-\beta)$, $1<\alpha\le 2$, and with the initial conditions (3.3) and (3.14).
 Then for the solution of (5.11), subject to the above conditions, there holds the formula

$$\eqalignno{N(x,t)&=\int_{-\infty}^{\infty}G(x-\xi,t)f(\xi){\rm d}\xi+\int_{-\infty}^{\infty}t~G_2(x-\xi,t)g(\xi){\rm d}\xi\cr
&+\mu\int_0^t(t-\tau)^{\alpha-1}[\int_{-\infty}^{\infty}G_{10}
(x-\tau,t-\tau)U(\xi,\tau){\rm d}\xi]{\rm d}\tau,&(5.12)\cr}
$$where the Green functions  $G(x,t)$, $G_2(x,t)$  and $G_{10}(x,t)$  are, respectively, given by (3.6), (3.16), and (5.4).}

\vskip.3cm
\noindent
Now, setting $\mu=1$ and $\theta=\beta=0$, we find that the Corollaries 8 and 9 give rise to the  expressions for the solution of  non-homogeneous one-dimensional fractional generalized wave equations as given by  Debnath [4, p.141].

\vskip.3cm\noindent
Finally, for $\alpha=1$ and $\theta=\phi=0$, Theorem 3 provides a solution of generalized diffusion equation with two space-fractional derivatives, which were recently studied by Pagnini and Mainardi [35].

\vskip.3cm\noindent\centerline{\bf References}

\vskip.3cm\noindent
[1]\hskip.3cm Bhatti, M. and Debnath, L.: On fractional Schr\"odinger and Dirac equations, {\it International Journal of Pure and Applied Mathematics}, {\bf 15} (2004), 1-11.

\vskip.2cm\noindent
[2]\hskip.3cm Chen, J., Liu, F., Turner, I., and Anh, V.: The fundamental and numerical solutions of the Riesz space-fractional reaction-dispersion equation, {\it The Australian and New Zealand Industrial and Applied Mathematics Journal (ANZIAM)}, {\bf 50} (2008), 45-57.

\vskip.2cm\noindent
[3]\hskip.3cm Cross, M.C. and Hohenberg, P.C.: Pattern formation outside of equilibrium, {\it Reviews of Modern Physics}, {\bf 65} (1993), 851-9128.

\vskip.2cm\noindent
[4]\hskip.3cm Debnath, L.: Fractional integral and fractional differential equations in fluid mechanics, {\it Fractional Calculus and Applied Analysis}, {\bf 6} (2003), 119-155.

\vskip.2cm\noindent
[5]\hskip.3cm Debnath, L.: {\it Nonlinear Partial Differential Equations for Scientists and Engineers}, Second Edition, Birkhäuser, Boston, Basel, Berlin 2005.

\vskip.2cm\noindent
[6]\hskip.3cm Diethelm, K.: {\it The Analysis of Fractional Differential Equations}, Springer, Berlin, Heidelberg, 2010.

\vskip.2cm\noindent
[7]\hskip.3cm Engler, H.: On the speed of spread for fractional reaction-diffusion {\it Interntional Journal of Differential Equations}, 2010, Article ID 315421, 16 pages.

\vskip.2cm\noindent
[8]\hskip.3cm Erd\'elyi, A., Magnus, W., Oberhettinger, F. and Tricomi, F.G.: {\it Higher Transcendental Functions}, Vol. {\bf 3}, McGraw-Hill, New York, Toronto 1955.

\vskip.2cm\noindent
[9]\hskip.3cm Feller, W.: On a generalization of Marcel Riesz' potentials and the semi-groups generated by them, {\it Meddelanden Lunds Universitets Matematiska Seminarium (Comm. Sém.Mathém. Université de Lund )}, {\bf Tome suppl. Dédié á M.Riesz, Lund}, (1952), 73-81.

\vskip.2cm\noindent
[10]\hskip.3cm Feller, W.: {\it An Introduction to Probability Theory and Its Applications}, Vol. {\bf 2}, Second Edition, Wiley, New York 1971 (First Edition, 1966).

\vskip.2cm\noindent
[11]\hskip.3cm Feynman, R.P. and Hibbs, A.R.: {\it Quantum Mechanics and Path Integrals}, McGraw-Hill, New York 1965.

\vskip.2cm\noindent
[12]\hskip.3cm Gafiychuk,V., Datsko, B., and Meleshko, V.: Mathematical modeling in pattern formation in sub and super-diffusive reaction-diffusion systems, 2006, arXiv:nlin/0811005.

\vskip.2cm\noindent
[13]\hskip.3cm Gafiychuk,V. Datsko, B., and Meleshko, V.: Nonlinear oscillations and stability domains in fractional reaction-diffusion systems, 2007, arXiv:nlin/0702013.

\vskip.2cm\noindent
[14]\hskip.3cm Gorenflo, R. and Mainardi, F.:  Approximation of Levy-Feller diffusion by random walk, {\it Journal for Analysis and its Applications}, {\bf 18} (1999), 1-16.

\vskip.2cm\noindent
[15]\hskip.3cm Guo, X. and Xu, M.: Some physical applications of Schr\"odinger equation, {\it Journal of Mathematical Physics}, 47082104 (2008): doi10, 1063/1.2235026 (9 pages).

\vskip.2cm\noindent
[16]\hskip.3cm Haubold, H.J., Mathai, A.M., and Saxena, R.K.:  Solutions of reaction-diffusion equations in terms of the H-function, {\it Bulletin  of the Astronomical Society, India}, {\bf 35} (2007), 681-689.

\vskip.2cm\noindent
[17]\hskip.3cm Haubold, H.J., Mathai, A.M., and Saxena, R.K.: Further solutions of reaction-dffusion equations in terms of the H-function, {\it Journal of Computational and Applied Mathematics}, {\bf 235} (2011), 1311-1316.

\vskip.2cm\noindent
[18]\hskip.3cm Henry, B.I. and Wearne, S.: Fractional reaction-diffusion, {\it Physica A}, {\bf 276} (2000), 448-455.

\vskip.2cm\noindent
[19]\hskip.3cm Henry, B.I. and Wearne, S.:  Existence of Turing instabilities in a two species reaction-diffusion system, {\it SIAM Journal of Applied Mathematics}, {\bf 62} (2002), 870-887.

\vskip.2cm\noindent
[20]\hskip.3cm Henry, B.I., Langlands, T.A.M., and Wearne, S.L.: Turing pattern formation in fractional activator-inhibitor systems, {\it Physical Review E} {\bf 72} (2005), 026101, 14 pages.

\vskip.2cm\noindent
[21]\hskip.3cm Kilbas, A.A., Srivastava, H.M., and Trujillo, J.J.: {\it Theory and Applications of Fractional Differential Equations}, Elsevier, Amsterdam 2006.

\vskip.2cm\noindent
[22]\hskip.3cm Kilbas, A,A. and Saigo, M.: {\it H-Transforms: Theory and Applications}, CRC Press, New York 2004.

\vskip.2cm\noindent
[23]\hskip.3cm Laskin, N.: Fractals and quantum mechanics, {\it Chaos}, {\bf 10} (2000), 780-790.

\vskip.2cm\noindent
[24]\hskip.3cm Laskin, N.: Fractional quantum mechanics and Levy path integrals, {\it Physics Letters A}, {\bf 268} (2000), 298-305.

\vskip.2cm\noindent
[25]\hskip.3cm Laskin, N.:  Fractional Schr\"odinger equation, {\it Physical Review E}, {\bf 66}, 056108, 2002.

\vskip.2cm\noindent
[26]\hskip.3cm Mainardi, F.: {\it Fractional Calculus and Waves in Linear Viscoelacsticity}, Imperial College Press, London 2010.

\vskip.2cm\noindent
[27]\hskip.3cm Mainardi, F., Luchko, Y., and Pagnini, G.: The fundamental solution of the space-time fractional diffusion equation, {\it Fractional Calculus and Applied Analysis}, {\bf 4} (2001), 153-102.

\vskip.2cm\noindent
[28]\hskip.3cm Mainardi, F., Pagnini, G., and Saxena, R.K.: Fox H-functions in fractional diffusion, {\it Journal of Computational and Applied Mathematics}, {\bf 178} (2005), 321-331.

\vskip.2cm\noindent
[29]\hskip.3cm Mathai, A.M., Saxena, R.K., and Haubold, H.J.: {\it The H-function: Theory and Applications}, Springer, New York 2010.

\vskip.2cm\noindent
[30]\hskip.3cm Metzler, R. and Klafter, J.: The random walk's guide to anomalous diffusion: A fractional dynamics approach, {\it Physics Reports}, {\bf 339} (2000), 1-77.

\vskip.2cm\noindent
[31]\hskip.3cm Miller, K.S. and Ross, B.: {\it An Introduction to the Fractional Calculus and Fractional Differential Equations}, Wiley, New York 1993.

\vskip.2cm\noindent
[32]\hskip.3cm Naber, M.: Distributed order fractional sub-diffusion, {\it Fractals}, {\bf 12} (2004), 23-32.

\vskip.2cm\noindent
[33]\hskip.3cm Nikolova, Y. and  Boyadjiev, L.: Integral transform methods to solve a time-space fractional  diffusion equation, {\it Fractional Calculus and Applied Analysis}, {\bf 13} (2010), 57-67.

\vskip.2cm\noindent
[34]\hskip.3cm Oldham, K.B. and Spanier, J.: {\it The Fractional Calculus Theory and Applications of Differentiation and Integration to Arbitrary Order}, Academic Press, New York 1974.

\vskip.2cm\noindent
[35]\hskip.3cm Pagnini, R. and Mainardi, F.: Evolution equations for a probabilistic generalization of Voigt profile function. {\it Journal of Computation and Applied Mathematics}, {\bf 233} (2010), 1590-1595.

\vskip.2cm\noindent
[36]\hskip.3cm Podlubny, I.: {\it Fractional Differential Equations}, Academicö Press, New York 1999.

\vskip.2cm\noindent
[37]\hskip.3cm Samko, S.G., Kilbas, A.A., and Marichev, O.I.:  {\it Fractional Integrals and Derivatives:  Theory and Applications}, Gordon and Breach Science Publishing, Switzerland 1993.

\vskip.2cm\noindent
[38]\hskip.3cm Saxena, R.K., Saxena, R., and Kalla, S.L.:  Computational solution of a fractional generalization of Schr\"odinger equation occurring in quantum mechanics, {\it Applied Matheatics and Computation}, {\bf 216} (2010), 1412-1417.

\vskip.2cm\noindent
[39]\hskip.3cm Saxena, R.K., Mathai, A.M., and Haubold, H.J.: Fractional reaction-diffusion equations,  {\it Astrophysics and Space Science}, {\bf 305} (2006a), 289-296.

\vskip.2cm\noindent
[40]\hskip.3cm Tofight, A.:  Probability structure of  time-fractional Schr\"odinger equation, {\it Acta Physica Polonica A}, {\bf 116} (2009), 111-118.

\bye